\documentclass[twocolumn,fleqn]{jpsj3}
\usepackage{graphicx,bm,color}

\title{Coexistence of Antiferromagnetism and Superconductivity in Iron-Based Superconductors}

\author{Yasunori Matsui$^{1}\thanks{Present address: Osaka Regional Headquarters, Japan Meteorological Agency, Osaka 540-0008, Japan}$
Takao Morinari$^{2}$, and Takami Tohyama$^{1,3}\thanks{E-mail address: tohyama@rs.tus.ac.jp}$ 
}
\inst{$^{1}$Yukawa Institute for Theoretical Physics, Kyoto University, Kyoto 606-8502, Japan \\
$^{2}$Graduate School of Human and Environmental Studies, Kyoto University, Kyoto 606-8501, Japan \\
$^{3}$Department of Applied Physics, Tokyo University of Science, Tokyo 125-8585, Japan
}
\recdate{April 28, 2014; accepted June 23, 2014}
\abst{We theoretically investigate the coexistence of antiferromagnetism and superconductivity in the iron-based superconductors by using the mean-field theory for two- and three-orbital models. We find that both the $s_{+-}$-wave and $s_{++}$-wave superconductivity can coexist with antiferromagnetism in the two models. On Dirac Fermi surfaces emerging in the antiferromagnetic phase, a superconducting-gap function has a node for $s_{++}$ wave but is nodeless for $s_{+-}$ wave. On the other hand, the gap function on non-Dirac Fermi surfaces is either nodeless or accidentally nodal, depending on the parameters of pairing interaction, which is independent of pairing symmetry.
}

\begin{document}
\sloppy
\maketitle

\def\Jonsc{J^{\mathrm{on}}_{{\mathrm{sc}}}}
\def\Jintersc{J^{\mathrm{inter}}_{{\mathrm{sc}}}}
\def\Im{{{\mathit{i}}}}

\section{Introduction}
\label{Intro}
Understanding the phase diagram of iron-based superconductors is a key to clarify the physics of superconductivity of these systems. Parent compounds of iron-based superconductors show an antiferromagnetic (AFM) spin density wave (SDW) below a N\'{e}el temperature. With hole or electron doping, magnetization is suppressed and superconductivity emerges. A close relationship of AFM and superconducting (SC) phases indicates that the formation of a Cooper pair is mediated by spin fluctuation, which leads to $s_{+-}$-wave order~\cite{Kuroki,Mazin} where the sign of the gap function on hole Fermi surfaces (FSs) centered at the momentum $\mathbf{k}=(0,0)$ is opposite to that on electron FSs at $\mathbf{k}$=($\pi$,0) and (0,$\pi$). On the other hand, it has been proposed that $s_{++}$-wave order, where the two signs are the same, emerges when orbital fluctuation is responsible for superconductivity~\cite{Onari,Kontani}. $s_{+-}$ and $s_{++}$ are possible candidates of SC symmetry in iron-based superconductors.

Since the AFM and SC phases are next to each other, the boundary of the two phases may provide useful information on superconductivity of iron-based superconductors. Intriguingly, Ba$_{1-x}$K$_x$Fe$_2$As$_2$, Ba(Fe$_{1-x}$Co$_x$)$_2$As$_2$ and BaFe$_2$(As$_{1-x}$P$_x$)$_2$ show a microscopically coexisting phase of AFM and SC orders, supported by neutron diffraction~\cite{Pratt,Fernandes_Neutron}, X-ray diffraction~\cite{Pratt,Nandi,Avci,Wiesenmayer}, and NMR~\cite{Iye,Li} experiments.

Such a coexistence phase has theoretically been investigated~\cite{Parker,Vorontsov,Fernandes,Ghaemi,Maiti,Schmiedt}, where it is commonly assumed that FSs in a paramagnetic phase consists of a near-circular hole pocket centered at $\mathbf{k}=(0,0)$ and an elliptical electron pocket at $\mathbf{k}$=($\pi$,0) and (0,$\pi$). The AFM order with wave vector ${\bf Q}$=($\pi$,0) mixes the hole and electron dispersions and a SDW gap is open. Resulting reconstructed FS is completely different from the FS of original paramagnetic phase. In the coexistence phase, it has been pointed out that the SC-gap function on the newly reconstructed FS is nodeless or has accidental nodes in the $s_{+-}$-wave order~\cite{Parker,Maiti} while nodes appear in the $s_{++}$ wave~\cite{Parker}. However, some of the theoretical studies have ignored multi-orbital nature of the dispersions, which leads to Dirac-type dispersions near the chemical potential in the AFM phase~\cite{Ran,Morinari}. Experimentally, an angular-resolved photoemission spectroscopy experiment has advocated the existence of the Dirac dispersions in the AFM phase~\cite{Richard}.

In this paper, taking into account multi-orbital properties through two- and three-orbital models, we investigate the coexistence of AFM order and the $s_{++}$- or $s_{+-}$-wave SC order by mean-field theory. We find that both the $s_{+-}$ wave and $s_{++}$ wave can coexist with antiferromagnetism in the two models. On the Dirac FSs, the SC-gap function has a node for $s_{++}$ but is nodeless for $s_{+-}$. The presence of node in $s_{++}$ is consistent with a previous theoretical work~\cite{Parker}. We also find that the gap function on non-Dirac FSs in the three-orbital model is either nodeless or nodal, depending on the parameters of pairing interaction. This behavior is independent of pairing symmetry.

The rest of this paper is organized as follows.  We introduce two-orbital and three-orbital models describing iron-based superconductors in \S~\ref{Model}.  In \S~\ref{Result}, phase diagrams and SC-gap functions in the coexisting phase are shown emphasizing some similarity and difference between the $s_{+-}$ wave and $s_{++}$ wave. The temperature dependence of order parameters in the coexisting phase is also discussed. The summary is given in \S~\ref{Summary}.

\section{Model}
\label{Model}
We consider multi-orbital tight-binding models for Fe3$d$ electrons on a square lattice,
\begin{equation}
H_{0}=\sum_{{\bm k},\sigma,\alpha,\beta}( T^{\alpha,\beta}({\bm k})-\mu \delta_{\alpha\beta} ) d^{\dagger}_{{\bm k},\alpha,\sigma} d_{{\bm k},\beta,\sigma},
\end{equation}
where $d^{\dagger}_{{\bm k},\alpha,\sigma}$ is the creation operator for a spin-$\sigma$ electron of momentum ${\bm k}$ and orbital $\alpha$. $\alpha$=1 and 2 for $d_{xz}$ and $d_{yz}$, respectively, in a two-orbital model, while $\alpha$=1, 2, and 3 for $d_{xz}$, $d_{yz}$, and $d_{xy}$, respectively, in a three-orbital model. $T^{\alpha,\beta}({\bm k})$ consists of energy levels and hopping matrix elements between the $\alpha$ orbitals. $\mu$ is the chemical potential. The lattice constant is set to unity. In the case of the two-orbital model~\cite{Raghu}, we take $T^{11}=-2t^{(2)}_{1}\cos k_x -2t^{(2)}_{2}\cos k_y -4t^{(2)}_{3}\cos k_x\cos k_y$, $T^{22}=-2t^{(2)}_{2}\cos k_x -2t^{(2)}_{1}\cos k_y -4t^{(2)}_{3}\cos k_x\cos k_y$, and $T^{12}=T^{21}=-4t^{(2)}_{4}\sin k_x\sin k_y$, with $t^{(2)}_{1}=-1.0$~eV, $t^{(2)}_{2}=1.3$~eV, and $t^{(2)}_{3} =t^{(2)}_{4} =-0.85$~eV. The energy levels of the two orbitals are set to be zero.

In the case of the three-orbital model~\cite{Daghofer}, we take $T^{11}=2t^{(3)}_{2}\cos k_x +2t^{(3)}_{1}\cos k_y +4t^{(3)}_{3}\cos k_x\cos k_y$, $T^{22}=2t^{(3)}_{1}\cos k_x +2t^{(3)}_{2}\cos k_y +4t^{(3)}_{3}\cos k_x\cos k_y$, $T^{33}=2t^{(3)}_{5}(\cos k_x+\cos k_y) +4t^{(3)}_{6}\cos k_x\cos k_y +\Delta_{{xy}}$, $T^{12}=T^{21}=4t^{(3)}_{4}\sin k_x\sin k_y$, $T^{13}=(T^{31})^* =2\Im t^{(3)}_{7}\sin k_x +4\Im t^{(3)}_{8}\sin k_x\cos k_y$, and $T^{23}=(T^{32})^* =2\Im t^{(3)}_{7}\sin k_y +4\Im t^{(3)}_{8}\sin k_y\cos k_x$, with $t^{(3)}_{1}=0.02$~eV, $t^{(3)}_{2}=0.06$~eV, $t^{(3)}_{3}=0.03$~eV, $t^{(3)}_{4}=-0.01$~eV, $t^{(3)}_{5}=0.2$~eV, $t^{(3)}_{6}=0.3$~eV, $t^{(3)}_{7}=-0.2$~eV, and $t^{(3)}_{8}=0.1$~eV. The energy level of $d_{xy}$ is $\Delta_{{xy}} =0.4$~eV. Both the two- and three-orbital models have been used for the study of the AFM and SC phases of iron-based superconductors~\cite{Ran,Raghu,Daghofer}.

In order to consider the AFM state, we adopt the following on-site Coulomb interactions:
\begin{align}
H_{\mathrm{AFM}} &= U\sum_i\sum_\alpha
n_{i,\alpha,\uparrow}
n_{i,\alpha,\downarrow}\notag \\
 &+
(U-2J)\sum_i\sum_{\alpha<\beta}\sum_{\sigma,\sigma^{\prime}}
n_{i,\alpha,\sigma}
n_{i,\beta,\sigma^{\prime}} \notag \\
 &-
J
\sum_i\sum_{\alpha<\beta}
\sum_{\sigma,\sigma^{\prime}}
d^{\dagger}_{i,\alpha,\sigma}
d_{i,\alpha,\sigma^{\prime}}
d^{\dagger}_{i,\beta,\sigma^{\prime}}
d_{i,\beta,\sigma}\notag \\
 & +
J
\sum_i\sum_{\alpha<\beta}
(
d^{\dagger}_{i,\alpha,\uparrow}
d^{\dagger}_{i,\alpha,\downarrow}
d_{i,\beta,\downarrow}
d_{i,\beta,\uparrow}
+
\mathrm{h.c.}
),
\end{align}
where the first and second terms are intra-orbital and inter-orbital Coulomb repulsions, respectively. The third term is Hund's coupling and the fourth term is inter-orbital pair hopping. $U$ and $J$ are the parameter of Coulomb and Hund's coupling interactions, respectively.

We solve a mean-field equation self-consistently with order parameter in the AFM state with ordering vector ${\bm Q}$, which is defined by
\begin{equation}
\langle n_{\bm{Q}\alpha\sigma} \rangle = \frac{1}{N}\sum_{\bm{k}} \langle d^\dagger_{\bm{k+Q},\alpha,\sigma} d_{\bm{k},\alpha,\sigma} \rangle,
\end{equation}
where $N$ is the number of momentum points in the first Brillouin zone of the paramagnetic phase and we take ${\bm Q}$ = ($\pi$,0). The average $\langle \cdots \rangle$ is taken at zero temperature unless otherwise explicitly stated. Note that we neglect inter-orbital order parameter $\langle n_{\bm{Q}\alpha\beta\sigma} \rangle = N^{-1}\sum_{\bm{k}} \langle d^\dagger_{\bm{k+Q},\alpha,\sigma} d_{\bm{k},\beta,\sigma} \rangle$ $ ({{\alpha \neq \beta}})$, because these values are almost zero for the present orbital models~\cite{Brydon}.

\begin{figure}[t]
 \begin{minipage}{0.49\hsize}
  \begin{center}
   \includegraphics[width=40mm,height=45mm]{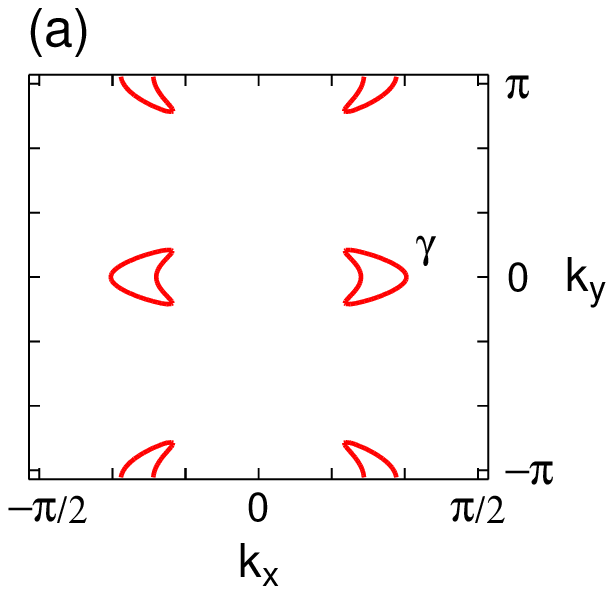}
  \end{center}
 \end{minipage}
 \begin{minipage}{0.49\hsize}
  \begin{center}
   \includegraphics[width=40mm,height=42mm]{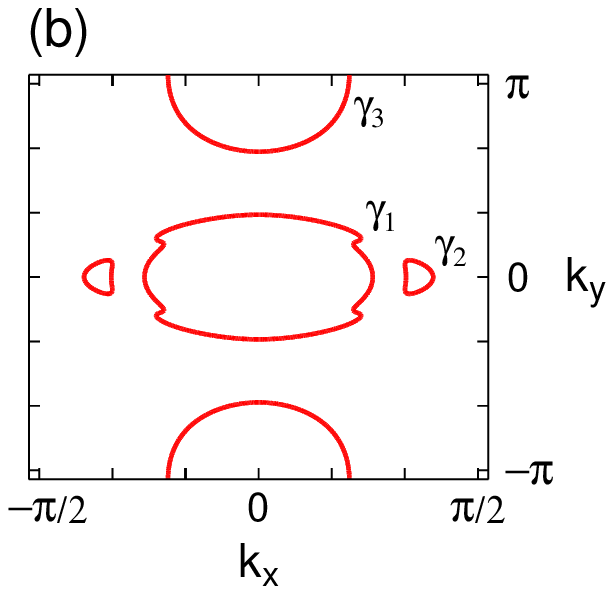}
  \end{center}
 \end{minipage}
\caption{(Color online) Fermi surfaces (FSs) in the AFM phase of (a) two- and (b) three-orbital models. We set the Coulomb interaction parameter $U/|t^{(2)}_{1}|= 3.0$ and electron density $n_{e}=2.0$ for (a) and $U/t^{(3)}_{1}= 32.5$ and $n_{e}=4.0$ for (b). $n_{e}$ is selected to be half-filling. The FSs are denoted by symbol $\gamma$ in (a) and $\gamma_{i}$ ($i$=1,2,and 3) in (b).  Note that the $\gamma$ and $\gamma_{2}$ FSs are associated with the Dirac FS whose node is located below the chemical potential.}
\label{Fig1}
\end{figure}

The FSs of the two- and three-orbital models in the AFM state at half filling are shown in Fig.~\ref{Fig1}. Here we set $J=0.25U$. There are small FSs, $\gamma$ in Fig.~\ref{Fig1}(a) and $\gamma_{2}$ in Fig.~\ref{Fig1}(b), associated with Dirac dispersions. The dispersions have a crossing point along $k_y=0$ and $k_y=\pi$ in the two-orbital model and along $k_y=0$ in the three-orbital model. This means that there is no SDW gap along the lines~\cite{Ran,Morinari}. This is because  off-diagonal elements in the hopping terms are zero along the lines and each orbital does not mix. The Dirac FSs characterize the electronic states of AFM state in iron-arsenide systems and cannot be described by a multi-band model where orbital components are ignored.

We consider two SC orders: $s_{++}$-wave order and $s_{+-}$-wave order. We assume that the pairing of $s_{++}$ comes from on-site pair with the same orbital. The pairing interaction then reads
\begin{equation}
H^{\mathrm{on}}_{\mathrm{sc}} = \Jonsc \sum_{i,\alpha<\beta}(
d^{\dagger}_{i,\alpha,\uparrow}
d^{\dagger}_{i,\alpha,\downarrow}
d_{i,\beta,\downarrow}
d_{i,\beta,\uparrow}
+\mathrm{h.c.}),
\label{Hon}
\end{equation}
where $\Jonsc$ is a parameter of on-site pairing interaction. 
Here, we ignored the scattering process of pairs within the same orbital.  The process is important when we argue the pairing mechanism due to spin fluctuations.~\cite{Kuroki,Mazin} However, in our mean-field treatment the effect of fluctuations is not taken into account. Therefore, inter-orbital pair scattering may be enough for our arguments shown below. The $s_{++}$ order parameter is then defined by 
\begin{equation}
\Delta^{\mathrm{on}}_{\alpha}  = \frac{1}{N}\sum_{\bm{k}} \langle d_{\bm{-k},\alpha,\downarrow} d_{\bm{k},\alpha,\uparrow} \rangle.
\end{equation}

It is known that the $s_{+-}$-wave pairing is organized from inter-site pairings~\cite{kariyado}. We introduce a next-nearest-neighbor pair with the same orbital for simplicity. Similar to the $s_{++}$ case (eq.~(\ref{Hon})), the pairing Hamiltonian with next-nearest-neighbor pair $\langle i,j \rangle$ reads
\begin{align}
H^{\mathrm{inter}}_{\mathrm{sc}} &= 
\Jintersc \sum_{\langle i,j \rangle} \sum_{\alpha\neq\beta} \sum_{\sigma\neq\sigma'}
\big(
d^\dagger_{j,\beta,\sigma} d^\dagger_{i,\beta,\sigma'} \notag \\
 &\ \ \ \ \ \ \ \ \ \ +
d^\dagger_{i,\beta,\sigma} d^\dagger_{j,\beta,\sigma'} \big)
d_{j,\alpha,\sigma'} d_{i,\alpha,\sigma},
\end{align}
where $\Jintersc$ is a parameter of inter-site pairing interaction. We define the $s_{+-}$ order parameter by  
\begin{equation}
\Delta^\mathrm{inter}_{\alpha} = \frac{1}{N}\sum_{\boldsymbol{\delta}} \sum_{\bm{k}} 2\mathrm{cos}(\bm{k}\cdot\boldsymbol{\delta})\langle d_{\bm{-k}\alpha\downarrow} d_{\bm{k} \alpha\uparrow}\rangle,
\end{equation}
where $\boldsymbol{\delta}=(1,1)$ and $(-1,1)$.

\section{Results and Discussions}
\label{Result}

 \begin{figure}
 \begin{center}
 \begin{minipage}{0.49\hsize}
  \begin{center}
   \includegraphics[width=45mm,height=40mm]{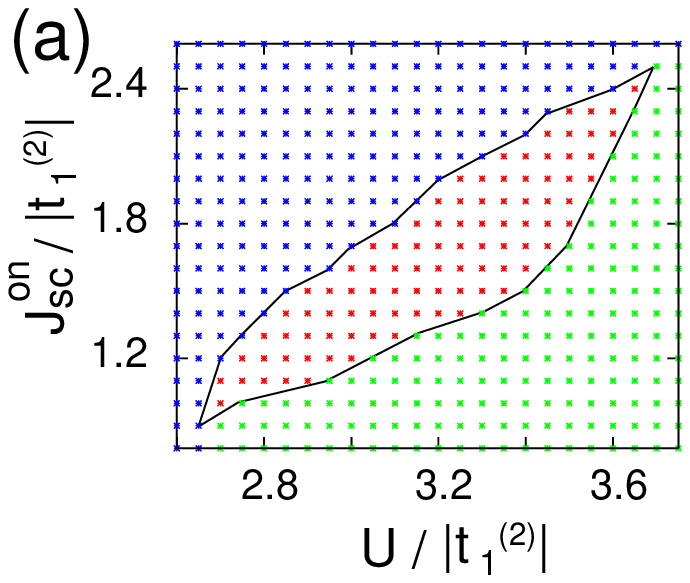}
  \end{center}
 \end{minipage}
 \begin{minipage}{0.49\hsize}
  \begin{center}
   \includegraphics[width=45mm,height=40mm]{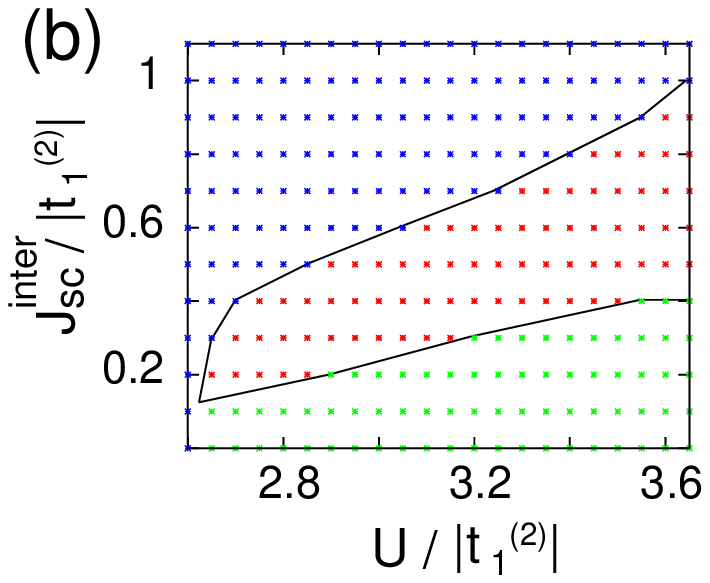}
  \end{center}
 \end{minipage}
  \begin{minipage}{0.49\hsize}
  \begin{center}
   \includegraphics[width=45mm,height=40mm]{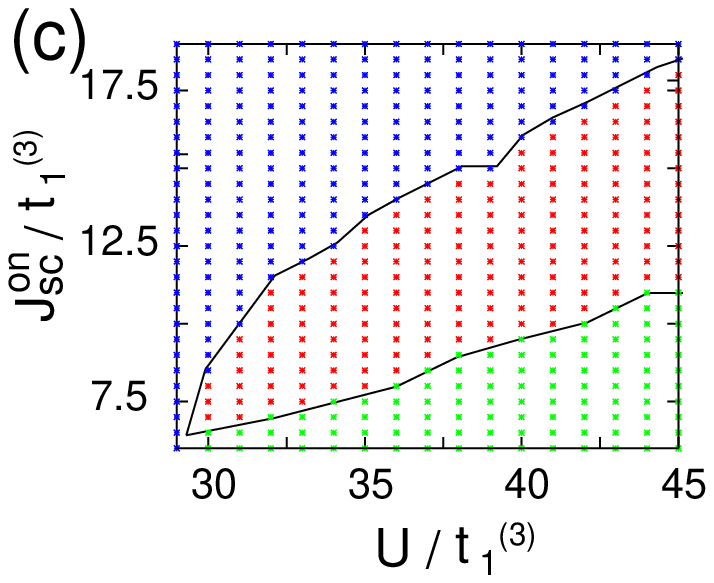}
  \end{center}
 \end{minipage}
 \begin{minipage}{0.49\hsize}
  \begin{center}
   \includegraphics[width=45mm,height=40mm]{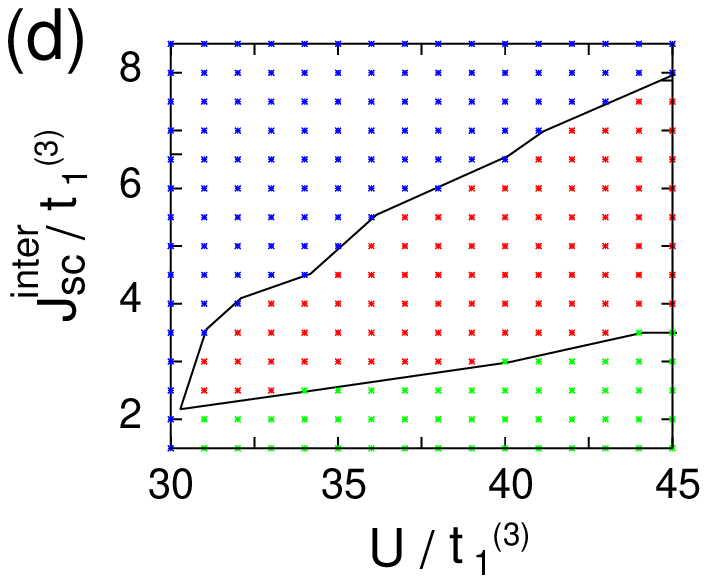}
  \end{center}
 \end{minipage}
 \end{center}
 \caption{(Color online) Phase diagram of the two-orbital model with $s_{++}$ wave (a) and with $s_{+-}$ wave (b) and of the three-orbital model with $s_{++}$ wave (c) and with $s_{+-}$ wave (d). The parameters for the two-orbital (three-orbital) model is scaled by $|t^{(2)}_{1}|$ ($t^{(3)}_{1}$). The area shown by blue dots (the dark-dotted area of small $U$ and large $\Jonsc$ and $\Jintersc$) represents SC phase, while the green-dotted area (light-dotted area of large $U$ and small $\Jonsc$ and $\Jintersc$) represents AFM phase. In between there is a coexisting phase represented by red dots. Solid lines representing phase boundary are of a guide to eyes. Electron density $n_{e}$ is fixed to 2.0 (4.0) for (a) and (b) ((c) and (d)).} 
\label{Fig2}
\end{figure}

The mean-field phase diagrams are shown in Fig.~\ref{Fig2}, where Figs.~\ref{Fig2}(a) and \ref{Fig2}(b) (Figs.~\ref{Fig2}(c) and \ref{Fig2}(d)) are for the two-orbital (three-orbital) model with $H=H_{0}+H_{{\mathrm{AFM}}}+H^{\mathrm{on}}_{{\mathrm{sc}}}$ ($H=H_{0}+H_{{\mathrm{AFM}}}+H^{\mathrm{inter}}_{{\mathrm{sc}}}$). There is a coexisting phase surrounded by the AFM and SC phases for all cases in the parameter region examined. In a previous study, a two-band model with symmetric hole and electron bands was found not to show such a coexistence for the $s_{++}$-wave paring~\cite{Fernandes}. However, we find that our two-orbital model exhibits a coexisting phase even for the $s_{++}$-wave symmetry as shown in Fig.~\ref{Fig2}(a). This indicates that a qualitatively different behavior emerges when we fully take the orbital degrees of freedom into account. The coexisting phase disappears for large $U$ and $\Jonsc$. This is in contrast with the $s_{+-}$ case shown in Fig.~\ref{Fig2}(b). The coexisting phase in the three-orbital model shows a similar $U$ dependence between $s_{++}$ in Fig.~\ref{Fig2}(c) and $s_{+-}$ in Fig.~\ref{Fig2}(d), which may be organized from non-Dirac FSs in the AFM state ($\gamma_1$ and $\gamma_3$ in Fig.~\ref{Fig1}(b)) as discussed below.

In order to understand the nature of SC order in the coexisting phase, we examine the SC gap function defined by
\begin{equation}
\langle
\gamma_{\epsilon\downarrow,\bm{-k}}
\gamma_{\epsilon\uparrow,\bm{k}}
\rangle
\equiv
\sum_{\alpha}
U_{\epsilon\alpha\downarrow}(\bm{-k})
U_{\epsilon\alpha\uparrow}(\bm{k})
\langle
d_{\bm{-k},\alpha,\downarrow}
d_{\bm{k},\alpha,\uparrow}
\rangle,
\label{Gap}
\end{equation}
where $U_{\epsilon\alpha\sigma}(\bm{k})$ is a unitary transformation from orbital $\alpha$ to band $\epsilon$ with spin $\sigma$ in the AFM state, and the basis for the average $\langle \cdots \rangle$ is given by $\sum_{\alpha}(d_{\bm{k},\alpha,\uparrow},d_{\bm{k+Q},\alpha,\uparrow},d^{\dagger}_{\bm{-k},\alpha,\downarrow},d^{\dagger}_{\bm{-k-Q},\alpha,\downarrow})$.

 \begin{figure}
 \begin{minipage}{0.3\hsize}
  \begin{center}
   \includegraphics[width=29mm,height=43mm]{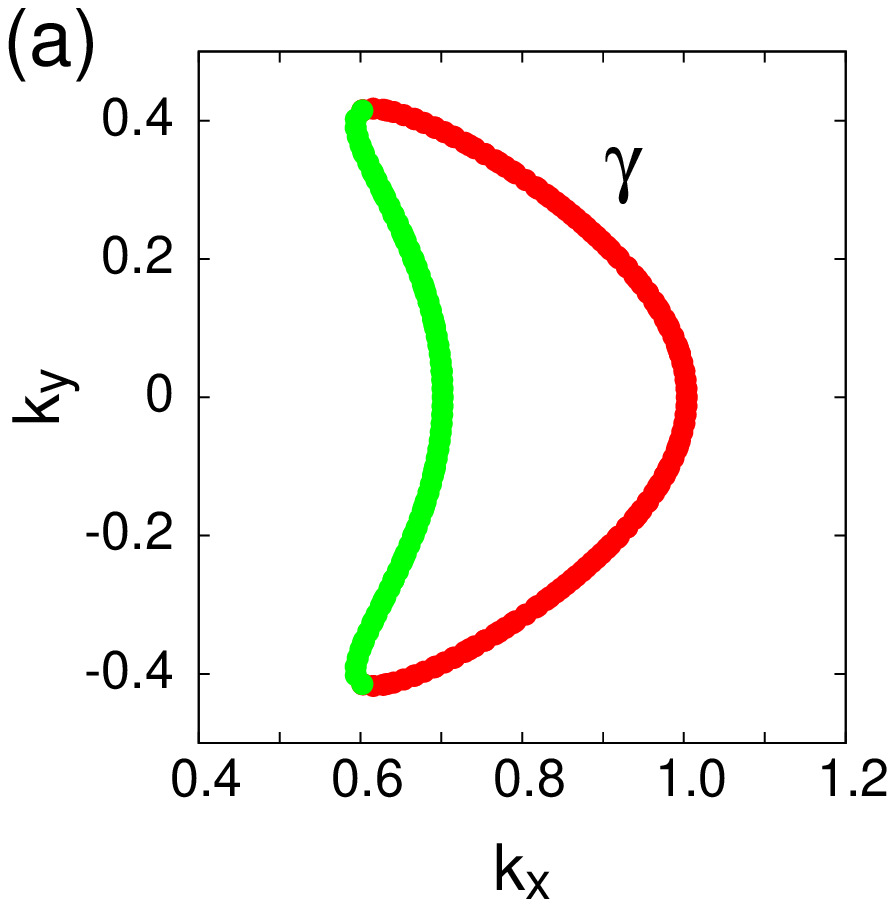}
  \end{center}
 \end{minipage}
 \begin{minipage}{0.3\hsize}
  \begin{center}
   \includegraphics[width=29mm,height=43mm]{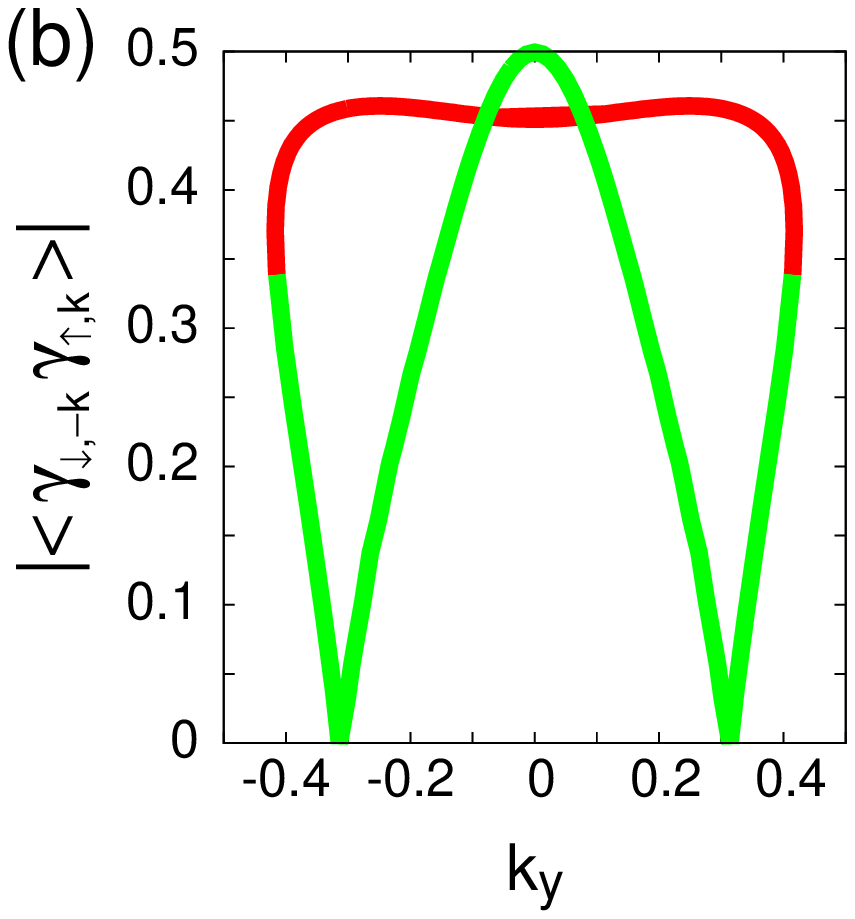}
  \end{center}
 \end{minipage}
  \begin{minipage}{0.3\hsize}
  \begin{center}
   \includegraphics[width=29mm,height=43mm]{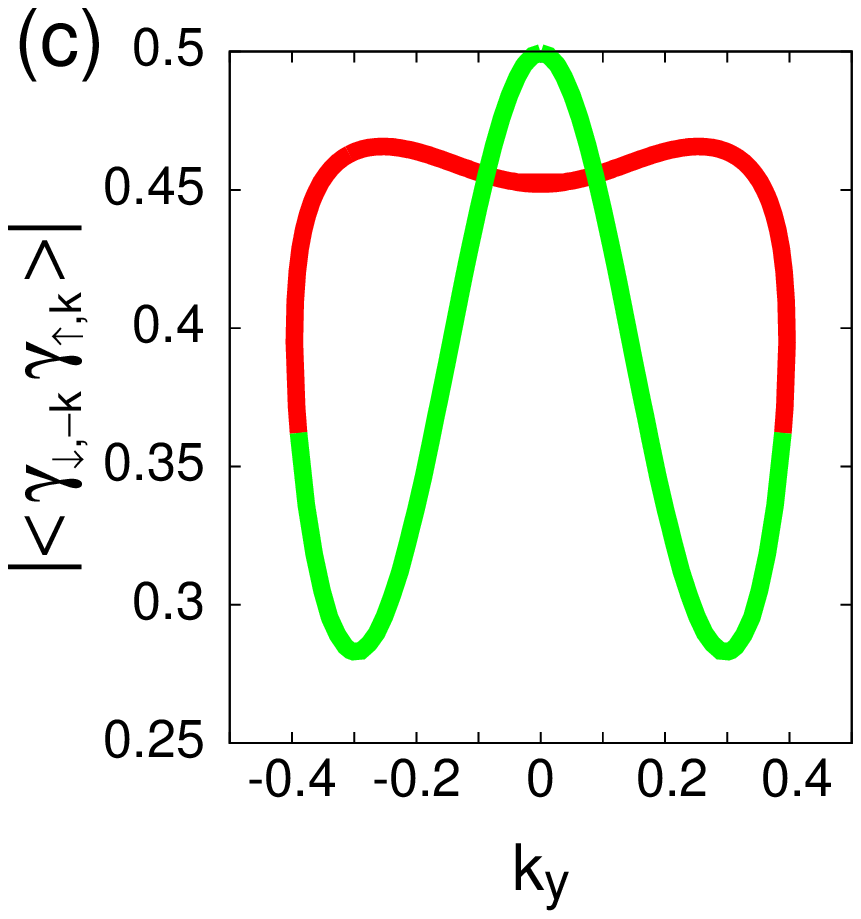}
  \end{center}
 \end{minipage}

 \begin{minipage}{0.3\hsize}
  \begin{center}
   \includegraphics[width=29mm,height=43mm]{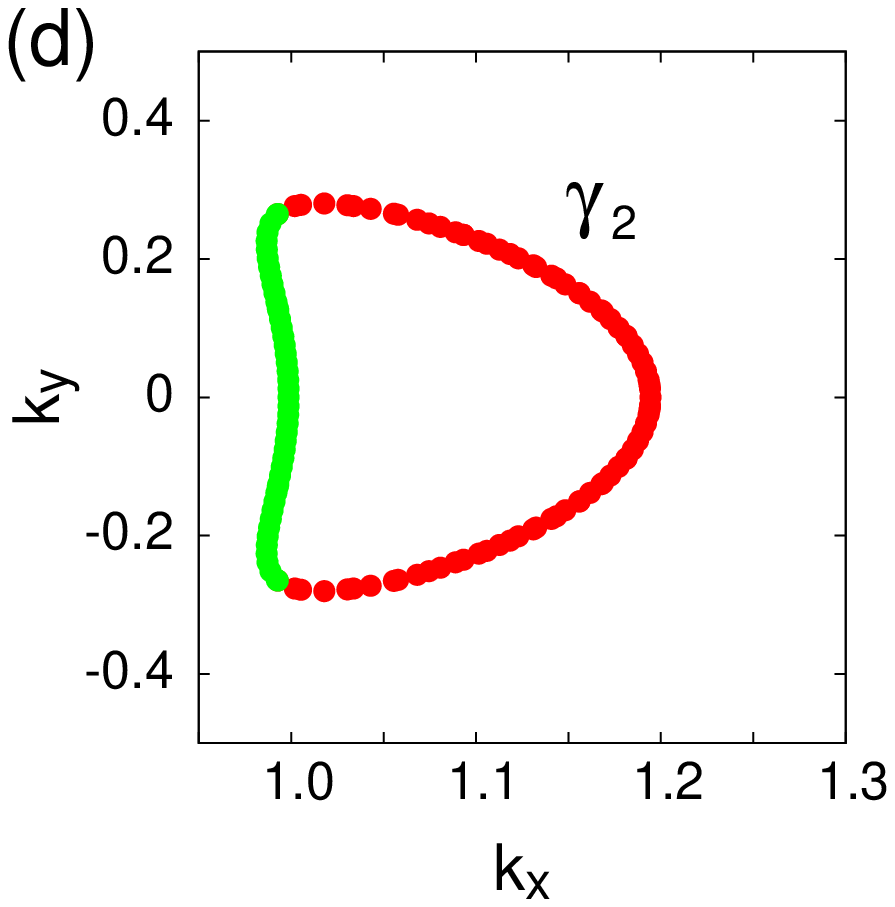}
  \end{center}
 \end{minipage}
 \begin{minipage}{0.3\hsize}
  \begin{center}
   \includegraphics[width=29mm,height=43mm]{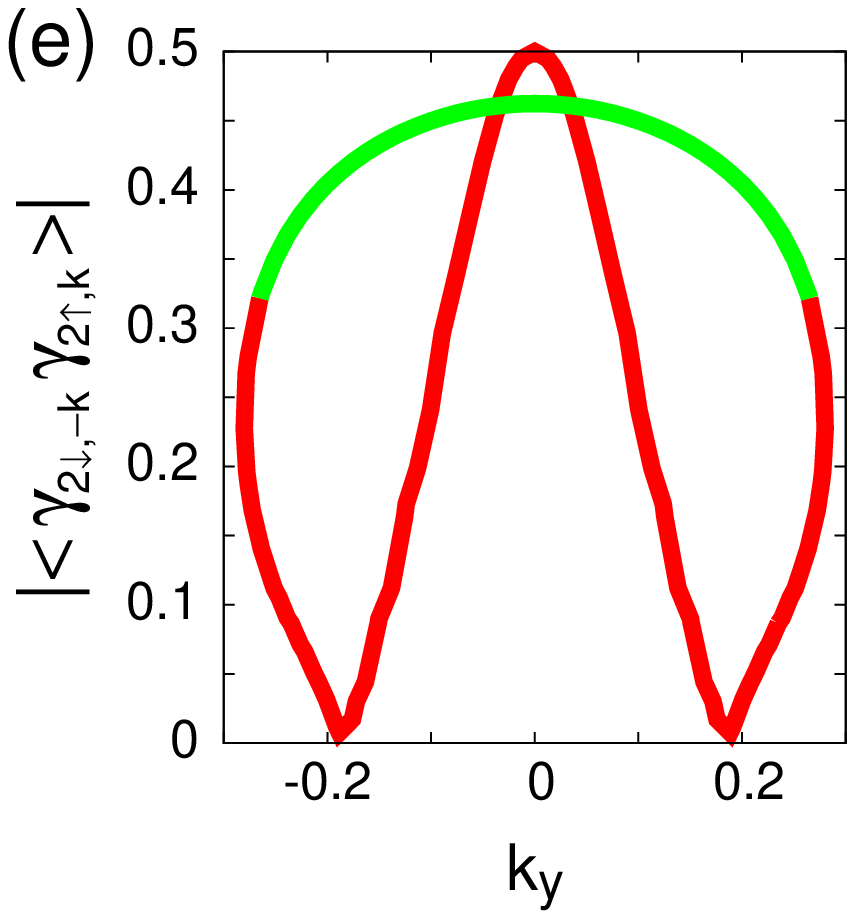}
  \end{center}
 \end{minipage}
  \begin{minipage}{0.3\hsize}
  \begin{center}
   \includegraphics[width=29mm,height=43mm]{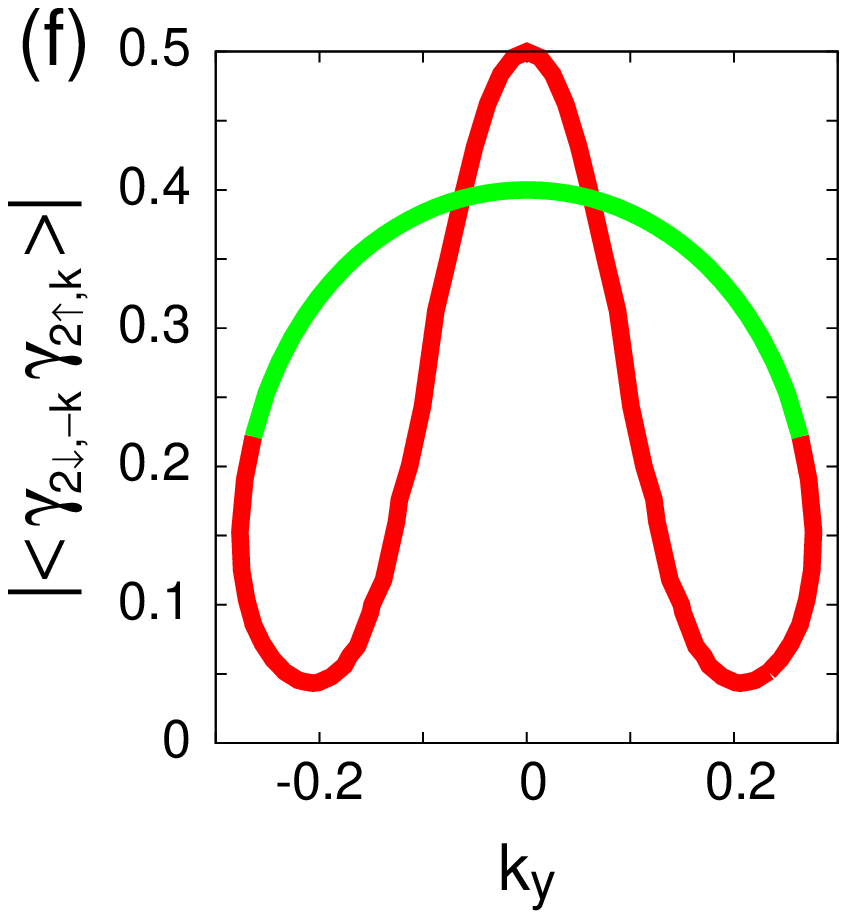}
  \end{center}
 \end{minipage}
\caption{(Color online) SC-gap functions on the Dirac-type FSs in a coexisting phase. (a) The Dirac FS denoted $\gamma$ in Fig.~\ref{Fig1}(a) in the two-orbital model, (b) SC-gap function on the FS for $s_{++}$, and (c) SC-gap function for $s_{+-}$. (d) The Dirac FS denoted $\gamma_2$ in Fig.~\ref{Fig1}(b) in the three-orbital model, (e) SC-gap function on the FS for $s_{++}$, and (f) SC-gap function for $s_{+-}$. The gap functions denoted by green (red) segments represent gap on the FS with green (red) segment in (a) and (d). The gap functions of $s_{++}$ wave in (b) and (e) show nodes, while those of $s_{+-}$ wave in (c) and (f) are fully gapped. The parameter values are $U/|t^{(2)}_{1}|= 3.0$, $\Jonsc/|t^{(2)}_{1}|= 1.6$, and $n_{e}=2.0$ for (b), $U/|t^{(2)}_{1}|= 3.0$, $\Jintersc/|t^{(2)}_{1}|= 0.5$, and $n_{e}=2.0$ for (c), $U/t^{(3)}_{1}= 32.5$, $\Jonsc/t^{(3)}_{1}= 9.0$, and $n_{e}=4.0$ for (e), $U/t^{(3)}_{1}= 32.5$, $\Jintersc/t^{(3)}_{1}= 3.5$, and $n_{e}=4.0$ for (f).
 }
\label{Fig3}
\end{figure}

Figure~3 demonstrates the SC gap along the Dirac-type FS shown in Figs.~\ref{Fig3}(a) and \ref{Fig3}(d) for the two-orbital and three-orbital models, respectively. The green (red) lines in Figs.~\ref{Fig3}(b) and \ref{Fig3}(c) correspond to the absolute value of gap function along the green (red) segment of Dirac FS in Fig.~\ref{Fig3}(a). Similarly, the green (red) lines in Figs.~\ref{Fig3}(e) and \ref{Fig3}(f) correspond to the green (red) segment in Fig.~\ref{Fig3}(d). In the case of the $s_{++}$ wave (Figs.~\ref{Fig3}(b) and \ref{Fig3}(e)) there are nodes on the gap functions, while in the $s_{+-}$ wave (Figs.~\ref{Fig3}(c) and \ref{Fig3}(f)) the gap functions are fully gapped. 

The sign of $U_{\epsilon\alpha\sigma}({\bm k})$ for the AFM state changes under the exchange of spin $\sigma$, which is caused by the SDW order. In the coexisting state with the $s_{++}$ wave,
this sign change gives rise to a cancellation of phase in the gap functions, leading to a node on the FS~\cite{Parker}. On the other hand, such a cancellation does not occur in the $s_{+-}$ wave. Therefore, measuring the gap functions on the Dirac FSs might be a good method to judge gap symmetry.

 \begin{figure}
 \begin{center}
 \begin{minipage}{0.49\hsize}
  \begin{center}
   \includegraphics[width=42mm,height=40mm]{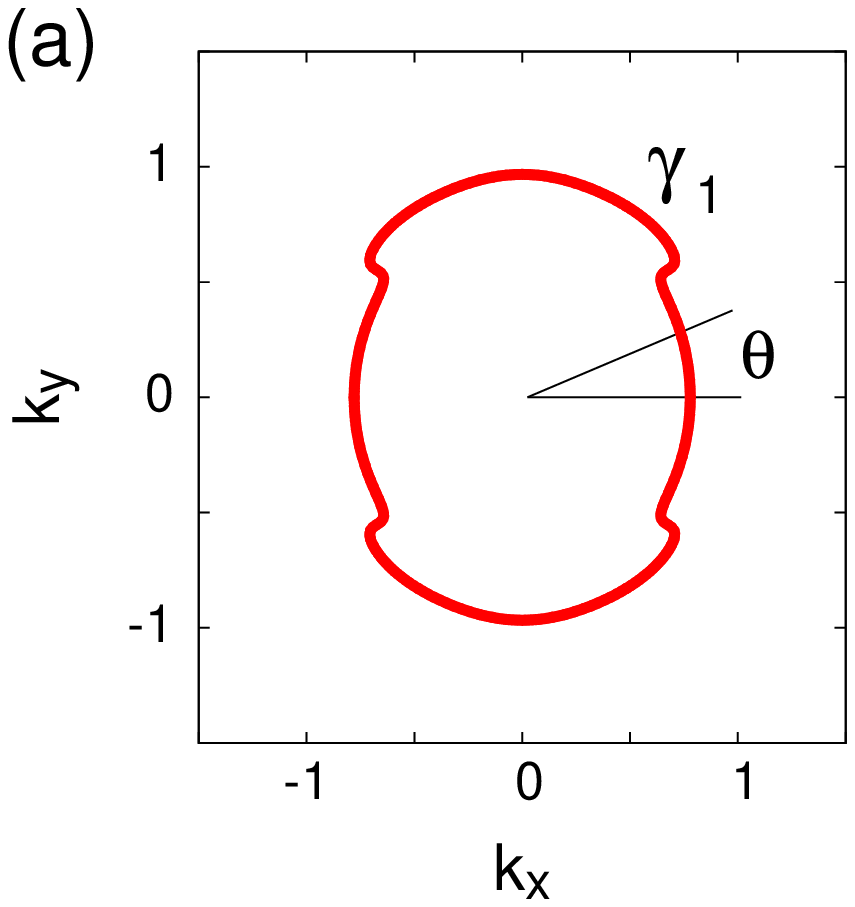}
  \end{center}
 \end{minipage}
 \begin{minipage}{0.49\hsize}
  \begin{center}
   \includegraphics[width=42mm,height=40mm]{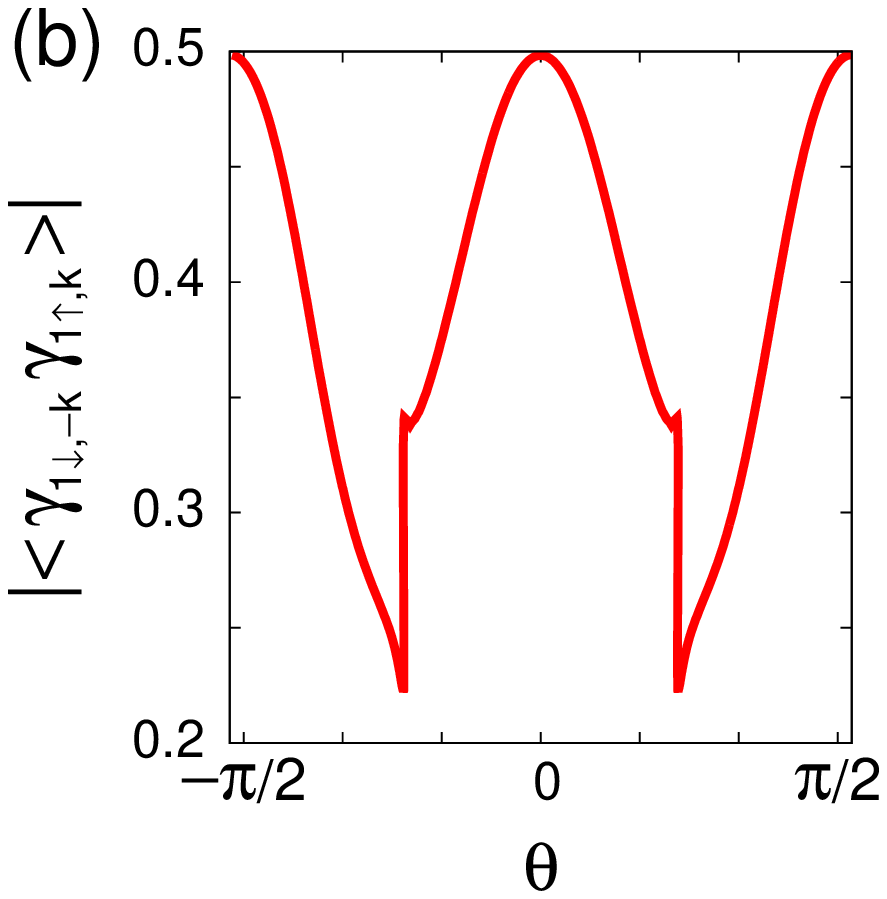}
  \end{center}
 \end{minipage}
  \begin{minipage}{0.49\hsize}
  \begin{center}
   \includegraphics[width=42mm,height=40mm]{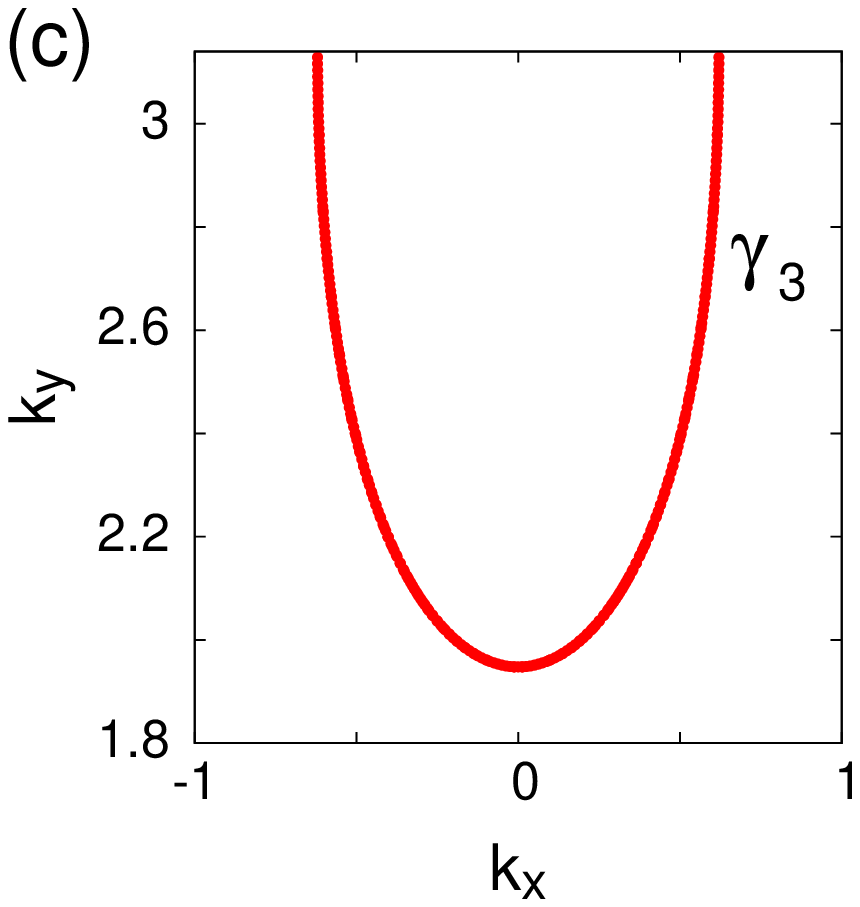}
  \end{center}
 \end{minipage}
 \begin{minipage}{0.49\hsize}
  \begin{center}
   \includegraphics[width=42mm,height=40mm]{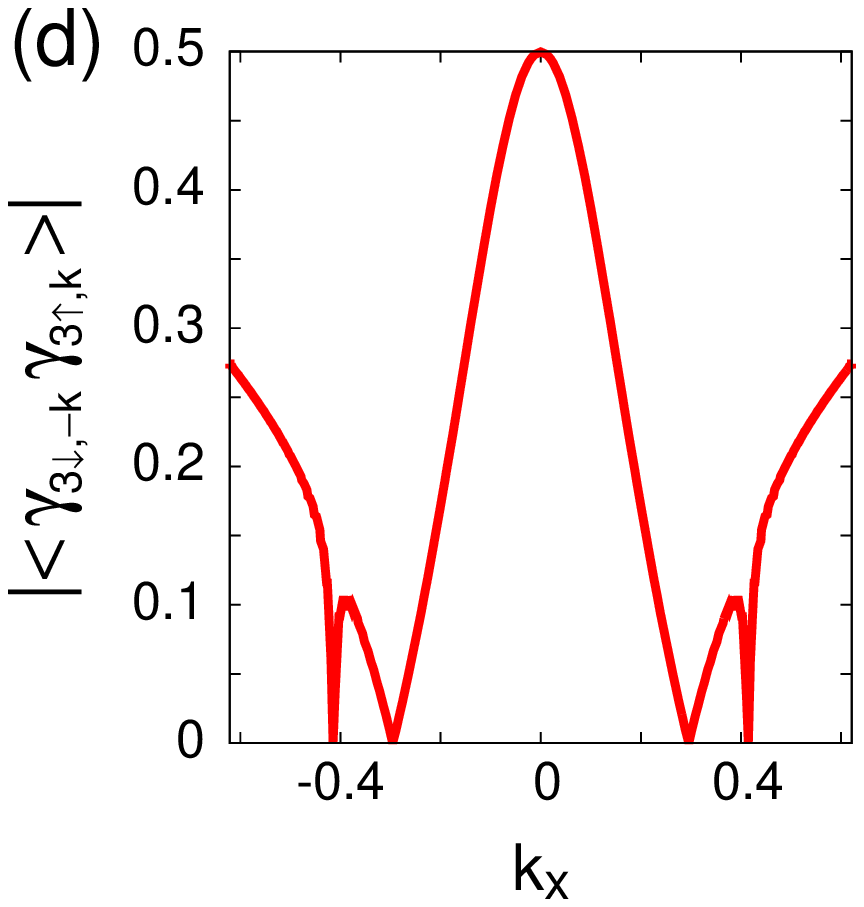}
  \end{center}
 \end{minipage}
 \end{center}
 \caption{(Color online) SC-gap function on non-Dirac-type FSs in a coexisting phase of $s_{++}$ wave for the three-orbital model. (a) FS denoted by $\gamma_1$ in Fig.~\ref{Fig1}(b). (b) SC-gap function on the $\gamma_1$ FS as a function of an angle $\theta$ measured from $k_y=0$. (c) $\gamma_3$ FS. (d) SC-gap function on the $\gamma_3$ FS. The parameter values are the same as Fig.~\ref{Fig3}(e). }
 \label{Fig4}
\end{figure}

In the three-orbital model, there are FSs unassociated with the Dirac dispersion, $\gamma_1$ and $\gamma_3$ in Fig.~\ref{Fig1}(b). Figure~\ref{Fig4} shows the gap function in a coexisting phase with $s_{++}$ wave for the three-orbital model. The gap function on the $\gamma_1$ FS is nodeless as shown in Fig.~\ref{Fig4}(b). On the other hand, the gap function on the $\gamma_3$ FS has nodes. Since the $\gamma_3$ FS is basically unaffected by the AFM order, the nodes are unrelated to the AFM order unlike the case of the Dirac FS. In this sense, the nodes are accidental. Actually it is possible to remove the nodes by varying the parameter $\Jonsc$. Note that similar gap functions are obtained for the $s_{+-}$ wave. 

\begin{figure}
 \begin{minipage}{0.49\hsize}
  \begin{center}
   \includegraphics[width=40mm,height=40mm]{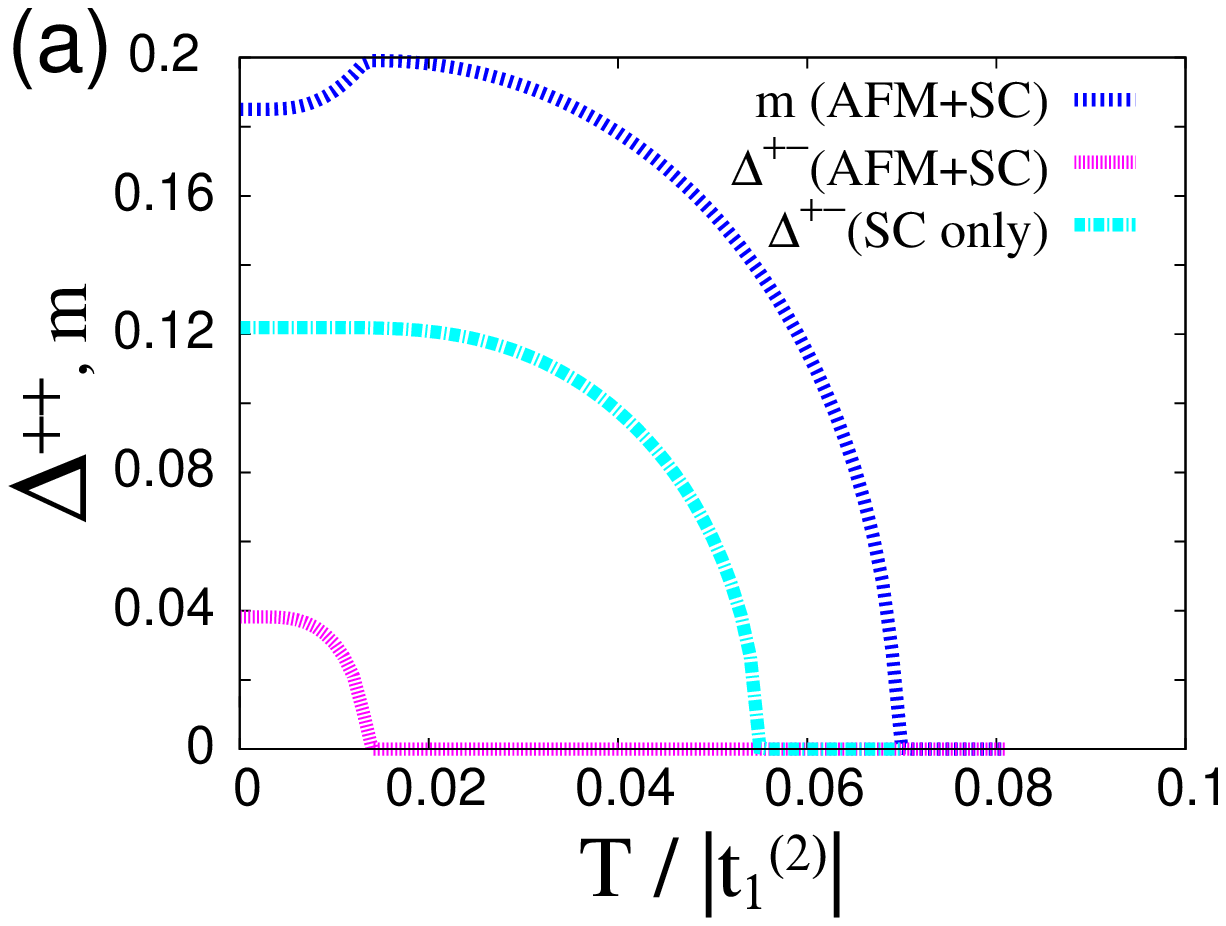}
  \end{center}
 \end{minipage}
 \begin{minipage}{0.49\hsize}
  \begin{center}
   \includegraphics[width=40mm,height=40mm]{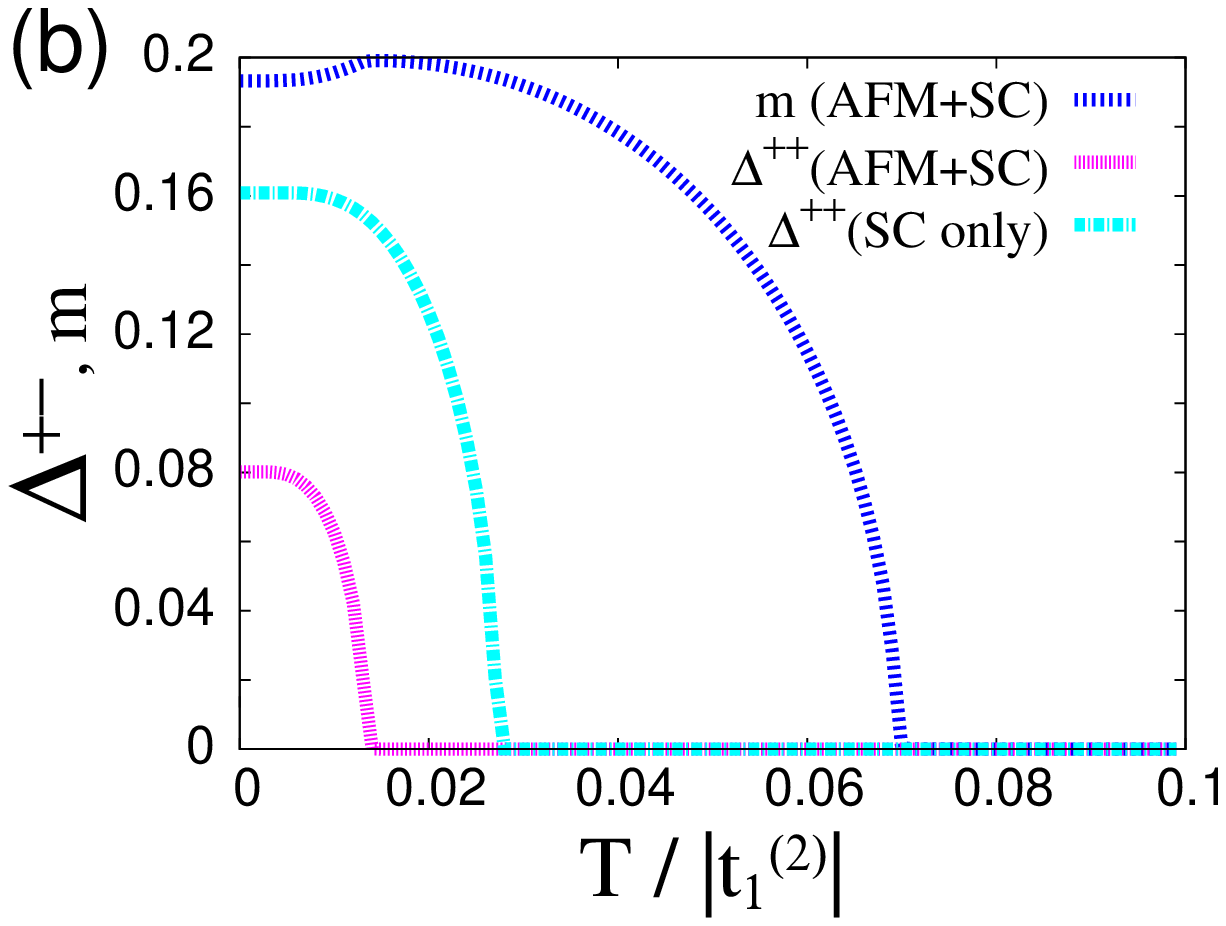}
  \end{center}
 \end{minipage}
  \begin{minipage}{0.49\hsize}
  \begin{center}
   \includegraphics[width=40mm,height=40mm]{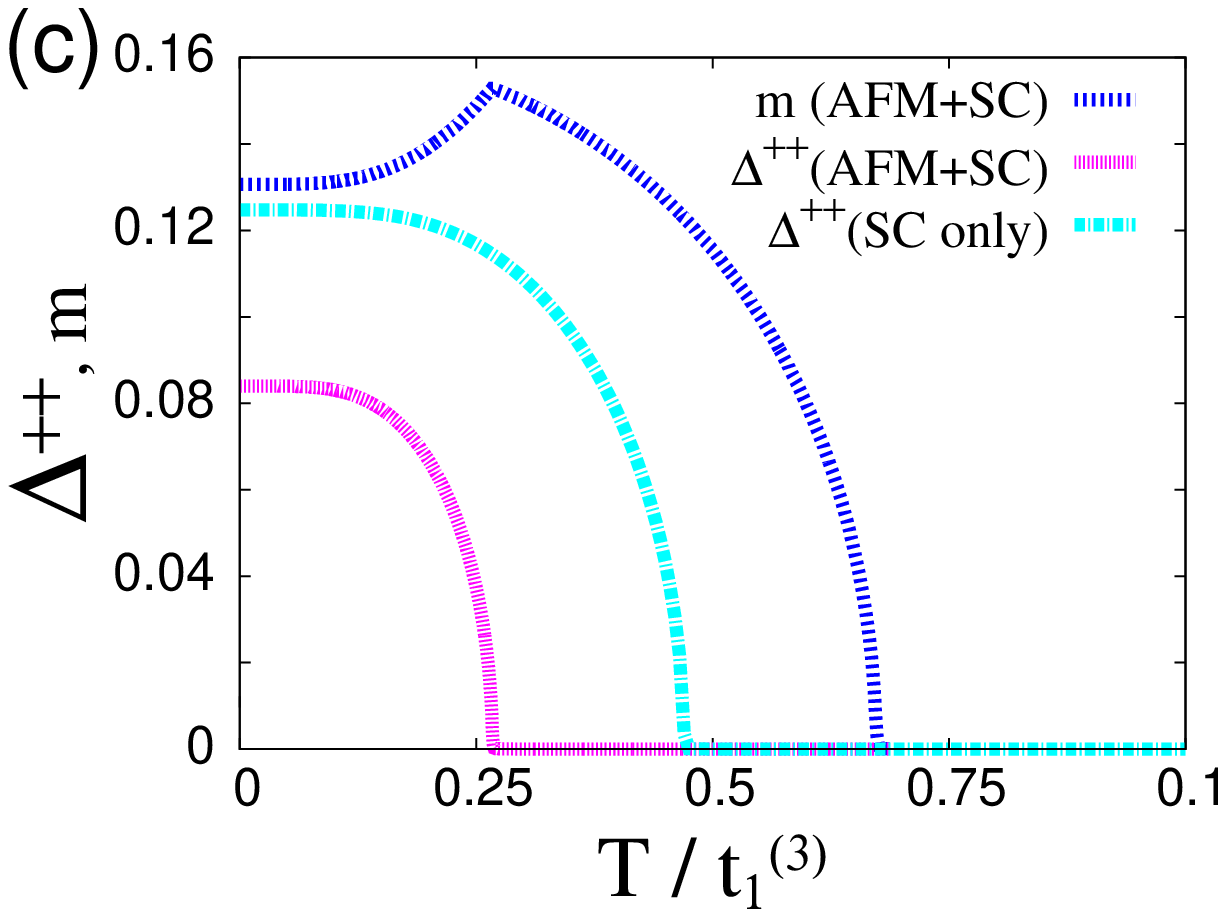}
  \end{center}
 \end{minipage}
 \begin{minipage}{0.49\hsize}
  \begin{center}
   \includegraphics[width=40mm,height=40mm]{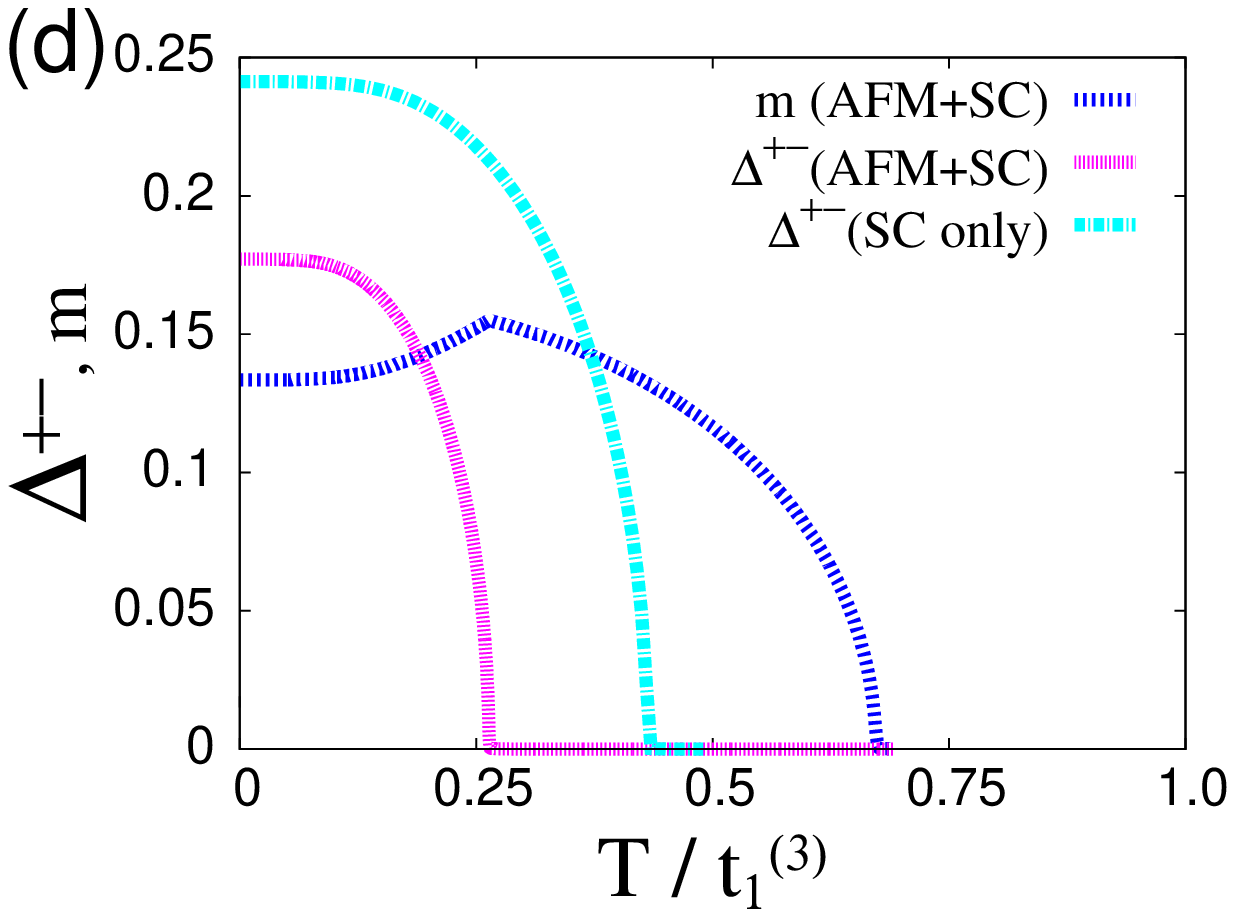}
  \end{center}
 \end{minipage}
 \caption{(Color online) The temperature dependence of SC order parameters ($\Delta^{++}$ for $s_{++}$ wave and $\Delta^{+-}$ for $s_{+-}$ wave) and AFM order parameter $m$. (a) $s_{++}$ and (b) $s_{+-}$ in the two-orbital model. (c) $s_{++}$ and (d) $s_{+-}$ in the three-orbital model. The dark blue lines with the highest transition temperature represent $m$ for the coexisting phase. The red lines with the lowest transition temperature represent SC order parameters for the coexisting phase, but the light blue lines between the dark blue and red lines are for the case without $U$ and $J$, i.e., only SC order. We take a parameter set for each panel: (a) $U/|t^{(2)}_{1}|= 3.0$, $\Jonsc/|t^{(2)}_{1}|= 1.6$, $n_{e}=2.0$, (b) $U/|t^{(2)}_{1}|= 3.0$, $\Jintersc/|t^{(2)}_{1}|= 0.42$, $n_{e}=2.0$, (c) $U/t^{(3)}_{1}= 35.0$, $\Jonsc/t^{(3)}_{1}= 12.5$, $n_{e}=4.0$, and (d) $U/t^{(3)}_{1}= 35.0$, $\Jintersc/t^{(3)}_{1}= 4.55$, $n_{e}=4.0$.
 }
 \label{Fig5}
\end{figure}

In order to clarify a competing behavior of AFM and SC states, we calculate the temperature dependence of AFM order parameter $m = \sum_\alpha (\langle n_{{\bm Q}\alpha\uparrow} \rangle - \langle n_{{\bm Q}\alpha\downarrow} \rangle)$ and SC order parameter $\Delta^{++} = \sum_\alpha  \Delta^\mathrm{on}_{\alpha}$ and $\Delta^{+-} = \sum_\alpha \Delta^\mathrm{inter}_{\alpha}$ for the $s_{++}$ and $s_{+-}$ waves, respectively. We take a typical parameter set describing coexisting phase for each case and the results are shown in Fig.~\ref{Fig5}. The dark blue lines correspond to $m$. In all cases, $m$ increases with decreasing temperature $T$, but at $T$ where SC order develops (red lines) $m$ decreases. This is a typical behavior of a coexisting phase observed theoretically and experimentally~\cite{Fernandes,Iye}. For comparison, SC order parameter without $m$ (equivalently $U=V=0$) is shown as a light blue line. In the two-orbital model, we find a large suppression of $\Delta^{++}$ under the presence of $m$ (Fig.~\ref{Fig5}(a)) as compared with $\Delta^{+-}$ (Fig.~5(b)). This is consistent with small coexisting region of $s_{++}$ shown in Fig.~\ref{Fig1}. The presence of nodes along the Dirac FS in $s_{++}$ may give less energy gain after superconductivity emerges. In the three-orbital model, the difference of the suppression of SC order parameter between $s_{++}$ (Fig.~\ref{Fig5}(c)) and $s_{+-}$ (Fig.~\ref{Fig5}(d)) is small. This implies that the non-Dirac $\gamma_{1}$ and $\gamma_{3}$ FSs, which are weakly related to AFM, are dominating the SC order and small contribution from the Dirac $\gamma_2$ FS.

In order to fully describe the electronic states of iron-pnictide superconductors, it is necessary to take a five-orbital model. In the present work, we mainly focus on the effect of the Dirac FS on the coexistence.  For this purpose, the most important factor is the presence of the Dirac FS, which is sufficiently achieved by the two- and three-orbital models. Our conclusions mentioned above will not change even if the number of orbitals is increased, as expected from the similarity between the two- and three-orbital models.

\section{Summary}
\label{Summary}
We have investigated the coexistence of superconducting and antiferromagnetic orders by mean-field calculations in the two- and three-orbital models. It has been known that there is no coexistence in the $s_{++}$ wave if one uses symmetric hole and electron dispersions without orbital components~\cite{Fernandes}. However, we have found that, if we take fully into account the orbital degrees of freedom, not only $s_{+-}$-wave superconductivity but also the $s_{++}$-wave superconductivity can coexist with antiferromagnetism. 

On Dirac-type Fermi surfaces constructed by SDW order, the gap functions are fully gapped in the $s_{+-}$ wave but have nodes in the $s_{++}$ wave. The presence of nodes in $s_{++}$ is in agreement with a previous study where orbital components were not taken into account~\cite{Parker}. On the non-Dirac Fermi surfaces in the three-orbital model, the gap functions are fully gapped or accidentally nodal, not strongly dependent on whether $s_{++}$ or $s_{+-}$. Therefore, it may be possible to distinguish pairing symmetry of iron-based superconductor by investigating the gap structure on Dirac Fermi surfaces by angle-resolved photo emission spectroscopy. 

\section*{Acknowledgment}
We acknowledge H. Kontani and S. Onari for useful discussions. This work was supported by Grant-in-Aid for Scientific Research from the Japan Society for the Promotion of Science, MEXT (Grant No.22740225).

\end{document}